\documentclass{article}
\usepackage{preprint}
\usepackage{amsmath,graphicx}
\usepackage{float}
\usepackage{listings}
\usepackage{mathtools}
\usepackage{upgreek}
\usepackage{tabularx}
\usepackage{siunitx}
\usepackage{caption,setspace}
\usepackage{enumitem}
\usepackage[numbers,sort&compress]{natbib}
\usepackage[toc,page]{appendix}
\usepackage[pagebackref,breaklinks,colorlinks,citecolor=cvprblue]{hyperref}
\usepackage[dvipsnames]{xcolor}

\renewcommand{\shorttitle}{M$^2$UGen}

\definecolor{cvprblue}{rgb}{0.21,0.49,0.74}
\usepackage[pagebackref,breaklinks,colorlinks,citecolor=cvprblue]{hyperref}

\title{M$^{2}$UGen: Multi-modal Music Understanding and Generation \\ with the Power of Large Language Models}
  
\author{Shansong Liu$^{1\dag*}$\\
ARC Lab, Tencent PCG\\
{\tt\small shansongliu@tencent.com}
\and
Atin Sakkeer Hussain$^{2*}$\\
National University of Singapore\\
{\tt\small atin.s@u.nus.edu}
\and
Qilong Wu$^{2*}$\\
National University of Singapore\\
{\tt\small qilong\_wu@u.nus.edu}
\AND
Chenshuo Sun$^{2}$\\
National University of Singapore\\
{\tt\small csun@nus.edu.sg}
\and
Ying Shan$^{1}$\\
ARC Lab, Tencent PCG\\
{\tt\small yingsshan@tencent.com}
}

\begin{document}
\twocolumn[{
\maketitle
\begin{center}
    \captionsetup{type=figure}
    \includegraphics[width=\textwidth]{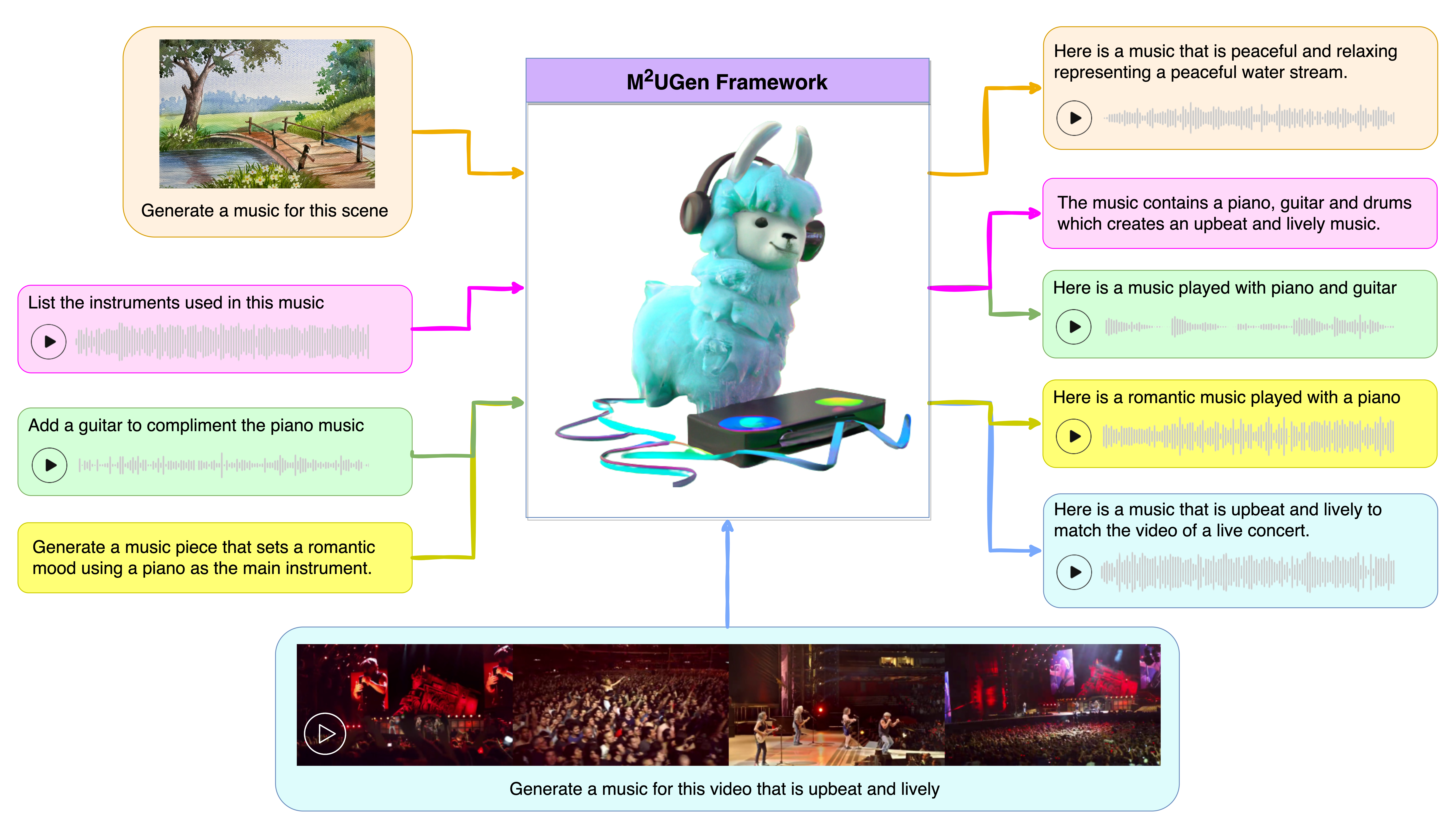}
    \captionof{figure}{Multi-modal music understanding and generation by our proposed M$^{2}$UGen framework.}
\end{center}
\vspace{4mm}
}]{\renewcommand*{\thefootnote}{$\dag$}\stepcounter{footnote}%
  \footnotetext{Corresponding author.}}
  {\renewcommand*{\thefootnote}{$*$}\stepcounter{footnote}%
  \footnotetext{Equal contribution.\\ \hspace*{1.2em} Please refer to and cite the latest version by searching for the title: ``MuMu-LLaMA: Multi-modal Music Understanding\\and Generation via Large Language Models''.}
  }

\begin{abstract}

The current landscape of research leveraging large language models (LLMs) is experiencing a surge. Many works harness the powerful reasoning capabilities of these models to comprehend various modalities, such as text, speech, images, videos, etc. They also utilize LLMs to understand human intention and generate desired outputs like images, videos and music. However, research that combines both understanding and generation using LLMs is still limited and in its nascent stage. To address this gap, we introduce a \textbf{M}ulti-modal \textbf{M}usic \textbf{U}nderstanding and \textbf{Gen}eration (\textbf{M$^{2}$UGen}) framework that integrates LLM's abilities to comprehend and generate music for different modalities. The M$^{2}$UGen framework is purpose-built to unlock creative potential from diverse sources of inspiration, encompassing music, image and video through the use of pretrained MERT, ViT, and ViViT models, respectively. To enable music generation, we explore the use of AudioLDM 2 and MusicGen. Bridging multi-modal understanding and music generation is accomplished through the integration of the LLaMA 2 model. Furthermore, we make use of the MU-LLaMA model to generate extensive datasets that support text/image/video-to-music generation, facilitating the training of our M$^{2}$UGen framework. We conduct a thorough evaluation of our proposed framework. The experimental results demonstrate that our model achieves or surpasses the performance of the current state-of-the-art models. 
\end{abstract}




\section{Introduction}
\label{sec:intro}

Multi-modal large language models (MLLMs) have emerged as a thriving area of research, captivating the current scholarly landscape \cite{yin2023survey}. They primarily serve as a bridge connecting diverse modalities, such as visual \cite{alayrac2022flamingo,li2023blip,pmlr-v209-xu23a}, audio \cite{tang2023salmonn,huang2023audiogpt,liu2023music}, 3D \cite{xu2023pointllm,sun20233d,wang2023chat} and so on, transcending mere textual interactions. This significant advancement greatly expands the application scenarios of large language models (LLMs).

Large language models are typically composed of a large number of parameters and trained on extensive datasets, endowing them with powerful comprehension and reasoning capabilities. Leveraging these qualities, researchers have utilized LLMs to achieve semantic understanding across various modalities. Examples include engaging in free-form conversations with humans \cite{openai_chatgpt,touvron2023llama}, comprehending audio/video events and performing event-based question answering \cite{tang2023salmonn,huang2023audiogpt,Maaz2023VideoChatGPT,zhao2023learning}, as well as captioning images/3D point cloud data \cite{chen2022visualgpt,li2023blip,xu2023pointllm,wang2023chat}. In addition to harnessing the capabilities of LLMs for multi-modal understanding, researchers have also strived to utilize these models to grasp the creative intentions of humans. For instance, they have explored generating images \cite{brade2023promptify}, videos \cite{hong2023cogvideo}, audio \cite{liu2023wavjourney}, or music \cite{copet2023simple} based on textual descriptions, thereby providing valuable assistance in artistic pursuits.

By integrating understanding and generation tasks within the framework of LLMs, we have the potential to significantly enhance the user experience. For example, users can leverage LLMs to summarize videos and generate accompanying audio commentary or suitable background music, thus assisting them in their video creation process. However, research that combines both understanding and generation using LLMs is still limited and in its nascent stage \cite{moon2022multimodal,ge2023planting,huang2023audiogpt,wu2023next,guo2023point,yang2023teal}. Among these few existing studies, NExT-GPT \cite{wu2023next} stands out as a notable advancement: it is a multi-modal large language model (MLLM) that excels in both understanding and generation tasks. NExT-GPT showcases several promising abilities, such as image/video question answering, text to image/video generation, audio understanding and generation, image/video generation for audio, and more. Despite these advancements, the exploration of music understanding and generation leveraging LLMs remains relatively unexplored. While NExT-GPT exhibits some capabilities in music understanding and generation, its proficiency in music-related tasks is modest due to the absence of specialized training on music datasets. To bridge this gap, we explore the use of LLMs for music understanding and multi-modal music generation in this work.

In the domain of music AI, significant progress has been made in developing tailored models for music understanding \cite{manco2021muscaps,doh2023lp,liu2023music,gardner2023llark} and multi-modal music generation \cite{di2021video,zhang2022vis2mus,zhuo2023video,agostinelli2023musiclm,copet2023simple,liu2023audioldm2}. For music understanding, the MU-LLaMA model \cite{liu2023music} stands as a representative, which is trained on a dedicated music question-answering dataset. It employs the MERT model \cite{li2023mert} as the music encoder and combines it with the LLaMA 2 model \cite{touvron2023llama}, demonstrating superior performance on music question answering and captioning tasks. Regarding multi-modal music generation, it can be divided into text-to-music generation, represented by models like MusicLM \cite{agostinelli2023musiclm}, MusicGen \cite{copet2023simple}, and AudioLDM 2 \cite{liu2023audioldm2}, as well as image/video-to-music generation, exemplified by models like Vis2Mus \cite{zhang2022vis2mus}, CMT \cite{di2021video} and V-MusProd \cite{zhuo2023video}. To the best of our knowledge, prior to the completion of this paper, there have been no other works capable of simultaneously encompassing music understanding and multi-modal music generation tasks using LLMs, except for the limited musical capabilities demonstrated by NExT-GPT. Therefore, in this work, we aim to contribute to this field by presenting our research findings.

In this work, we propose the \textbf{M}ulti-modal \textbf{M}usic \textbf{U}nderstanding and \textbf{Gen}eration (\textbf{M$^{2}$UGen}) framework capable of music understanding and drawing inspiration from multi-modal inputs to generate music. Our framework employs multiple modal encoders to represent image, video and music inputs, respectively. In line with the conventions of the computer vision field, we select ViT \cite{dosovitskiy2021image} and ViViT \cite{arnab2021vivit} as the encoders for the image and video modalities. The MERT model \cite{li2023mert}, which has exhibited outstanding performance in downstream music tagging tasks in the MU-LLaMA work \cite{liu2023music}, is chosen as our music encoder. The feature representations obtained from the input encoders of different modalities are then fed into their respective understanding adaptors. The LLaMA 2 model \cite{touvron2023llama} comprehends these modality signals and input intentions to carry out downstream tasks. For the music generation task, we explore and compare two music decoders, which are AudioLDM 2 \cite{liu2023audioldm2} and MusicGen \cite{copet2023simple}, while music understanding is directly addressed by the LLaMA 2 model.

In the realm of LLM-assisted music understanding and generation, there is a notable scarcity of readily available training data. The MusicCaps dataset \cite{agostinelli2023musiclm}, which stands as the largest publicly available dataset for text-to-music generation, only comprises approximately 28.52 hours of music accompanied by captions. Moreover, there is a dearth of datasets for the training of image/video-to-music generation. Hence, in order to tackle this data scarcity issue, we employ MU-LLaMA \cite{liu2023music} and MPT-7B \cite{MosaicML2023Introducing} models to generate diverse modality-music pairs for training our models. Furthermore, we will release our constructed datasets later to contribute to the development of the community.

Our contributions are summarized as follows:
\begingroup
\renewcommand\labelenumi{ \textbf{\theenumi)}}
\begin{enumerate}[leftmargin=.2in]
    \setlength{\itemsep}{2mm}
    \item We introduce the M$^{2}$UGen framework, an advancement capable of simultaneously encompassing music understanding and multi-modal music generation tasks, aiming to assist users in music related artistic creation.

    \item We propose a systematic approach for generating large multi-modal music oriented instruction datasets for the training of our M$^{2}$UGen model.

    \item We conduct comprehensive evaluations on various subtasks such as music question answering, text/image/video-to-music generation and music editing, showcasing performance levels that surpass or are on par with the state-of-the-art (SOTA) models.
    
\end{enumerate}

\section{Related Works}
\label{sec:related}

\begin{figure*}[t]
    \centering
    \includegraphics[width=\textwidth]{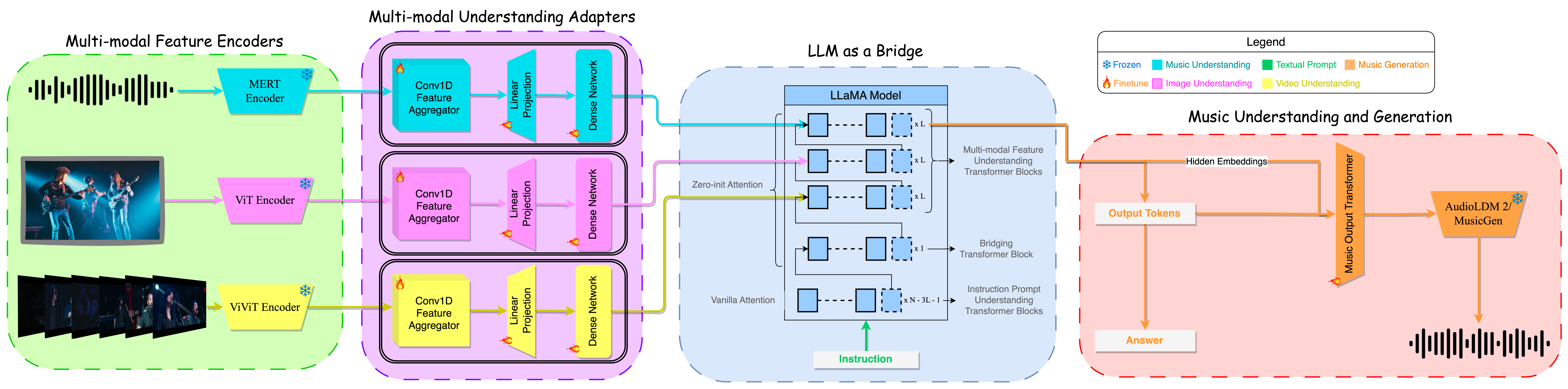}
    \caption{\textbf{Multi-modal Music Understanding and Generation Model (M$^{2}$UGen).} The model is divided into four parts (from left to right): 
(1) Pre-trained feature encoders to generate representations from music/images/videos; 
(2) Multi-modal understanding adapters to fuse the modality representations into the LLaMA 2 model; 
(3) The LLaMA 2 model that takes input from the adapters to learn context information; 
(4) The output projection layer that projects the LLaMA 2 model's output embeddings for the music decoder.}
  \label{fig:model}
\end{figure*}

\paragraph{Multi-modal understanding.} Our world is filled with diverse multi-modal information, while a myriad of AI models incessantly emerges, striving to comprehend various forms of information. The tasks themselves encompass a wide spectrum, ranging from audio/visual classification \cite{hershey2017cnn,dosovitskiy2021image,arnab2021vivit}, audio/visual question answering \cite{fayek2020temporal,antol2015vqa,lei-etal-2018-tvqa,yang2021justask}, audio/visual captioning \cite{Mei2021act,yao2017boosting,iashin2020multi} or tagging \cite{gong2021psla,zhang2016fast,ilyas2019deep}, audio/video event detection \cite{dinkel2021towards,li2019zero}, audio/video summarization \cite{vartakavi2021audio,ji2019video}, and so forth. These techniques have witnessed a rapid evolution over the past few decades, flourishing with remarkable vitality. In these research works, ViT \cite{dosovitskiy2021image} was a milestone in the field of computer vision, and many existing visual-related researches \cite{alayrac2022flamingo,li2022blip,zhai2022scaling} utilized ViT as the image encoder. ViViT \cite{arnab2021vivit}, which was inspired by ViT, incorporated temporal and spatial information to perform video representation. In our work, we adopt ViT and ViViT as encoders for image and video modalities, respectively. Regarding the music modality, the MU-LLaMA paper \cite{liu2023music} compared several SOTA models in their music feature representation section, including ImageBind \cite{girdhar2023imagebind}, Jukebox \cite{dhariwal2020jukebox}, MERT \cite{li2023mert}, and others. The authors revealed that the MERT encoder achieved the best performance in downstream music tagging tasks. Hence, following \cite{liu2023music}, we choose MERT as the music encoder in this work.

\vspace{-0.3cm}
\paragraph{Multi-modal Music Generation.} Research in music generation has made significant strides in recent years, particularly with the rise of Transformer \cite{vaswani2017attention} and diffusion \cite{ho2020denoising} models. Some notable works representing the utilization of text as input for guiding music generation include MusicLM \cite{agostinelli2023musiclm}, MusicGen \cite{copet2023simple}, Mo\^usai \cite{schneider2023mo}, Noise2music \cite{huang2023noise2music}, and AudioLDM 2 \cite{liu2023audioldm2}. Among these, AudioLDM 2 and MusicGen stand out for their exceptional music generation abilities. AudioLDM 2 is a general-purpose audio generation model based on the diffusion process, while MusicGen consists of a single autoregressive Transformer decoder. In the proposed M$^{2}$UGen framework, we explore and compare these two music decoders. There have been a few works in investigating music generation guided by alternative modalities, such as Vis2Mus \cite{zhang2022vis2mus} for generating music from images, and CMT \cite{di2021video} and V-MusPod \cite{zhuo2023video} for generating music from videos. However, these approaches are limited to a single modality as the instruction condition. In contrast, this paper presents a novel approach that integrates multiple modalities, including text, image, and video, leveraging the LLM framework to achieve multi-modal music generation.

\vspace{-0.3cm}
\paragraph{LLM-assisted Multi-modal Understanding and Generation.} MLLMs have emerged as a prominent research topic, with researchers tirelessly equipping these models with various functionalities. For example, Macaw-LLM \cite{lyu2023macaw} integrates text, images, videos, and audio into a unified framework, enabling multi-modal comprehension. DreamLLM \cite{dong2023dreamllm} and InternLM-XComposer \cite{zhang2023internlm} generate text and images in an interleaved manner, enhancing the multi-modal conversational abilities between users and LLMs. For instance, InternLM-XComposer can compose an article with accompanying illustrations. SEED-LLaMA \cite{ge2023making} leverages LLaMA \cite{touvron2023llama} and diffusion models for image understanding and question answering, as well as image generation. The NExT-GPT model \cite{wu2023next}, which is most relevant to our proposed framework, facilitates interleaved conversations involving text, images, videos, and audio. However, its limited music training data restricts its musical capabilities. In this paper, we introduce the M$^{2}$UGen framework, which enables music understanding and multi-modal music generation guided by multiple modalities. It also allows for modifying input music based on music prompts and text. We believe that our work will make a valuable contribution to the community of AI-assisted artistic creation.
\section{M$^{2}$UGen Model Architecture \& Training}

The architecture of the M$^{2}$UGen model is illustrated in Figure \ref{fig:model}. In this section, we provide an in-depth description of the M$^{2}$UGen architecture and elaborate on the training methods employed during the development of this model.

\subsection{Multi-modal Feature Encoders}


In order to accomplish multi-modal music understanding and generation, the M$^{2}$UGen model needs to be able to process multi-modal inputs. To achieve this, it incorporates frozen pre-trained encoders capable of comprehending various modalities, including music, images, and videos. These encoders extract relevant information to address tasks associated with music question answering and music generation within the M$^{2}$UGen framework. Specifically, the following encoders are employed for processing involved modalities:


\vspace{-0.3cm}
\paragraph{MERT Encoder} Notably, the MERT model \cite{li2023mert} has demonstrated exceptional performance in downstream music tagging tasks, as highlighted by Liu et al. (2023) \cite{liu2023music}, surpassing alternative models such as Wav2CLIP \cite{wu2022wav2clip}, ImageBind \cite{girdhar2023imagebind}, and Jukebox \cite{dhariwal2020jukebox}. Hence, we incorporate the MERT model as the music encoder to generate feature embeddings for music inputs. The shape of the output embedding is $(25, 1024)$, which is obtained by stacking the 24 hidden layers and the final output layer of the MERT model.


\vspace{-0.3cm}
\paragraph{ViT Encoder} Vision Transformer (ViT) \cite{dosovitskiy2021image} is a prominent breakthrough due to its performance and a prevailing image encoder in the field of computer vision. It splits an image into a series of fixed-sized patches and transforms them into patch embeddings, which are then fed into the Transformer encoder along with positional encodings. We adopt ViT as the encoder for image inputs, and it produces feature embeddings with a dimension of $(197, 768)$, where $197$ is the number of $16\times16$ patches in a $224\times224$ input image plus the final output layer, while $768$ corresponds to the hidden size of the Transformer.

\vspace{-0.3cm}
\paragraph{ViViT Encoder} The Video Vision Transformer (ViViT) model, as introduced by Arnab et al. (2021) \cite{arnab2021vivit}, represents one of the initial successful implementations of purely Transformer-based models for video comprehension. The ViViT model extracts spatio-temporal tokens from the input video and subsequently processes them through a sequence of Transformer layers to generate feature embeddings. The ViViT model produces embeddings with a shape of $(3137, 768)$, where $3137$ is derived from the total count of $16\times16$ patches sampled uniformly from 32 frames of size $224\times224$, including the final output layer, and $768$ is the hidden size of the Transformer.


\begin{figure}[htbp]
    \centering
    \includegraphics[width=0.9\columnwidth]{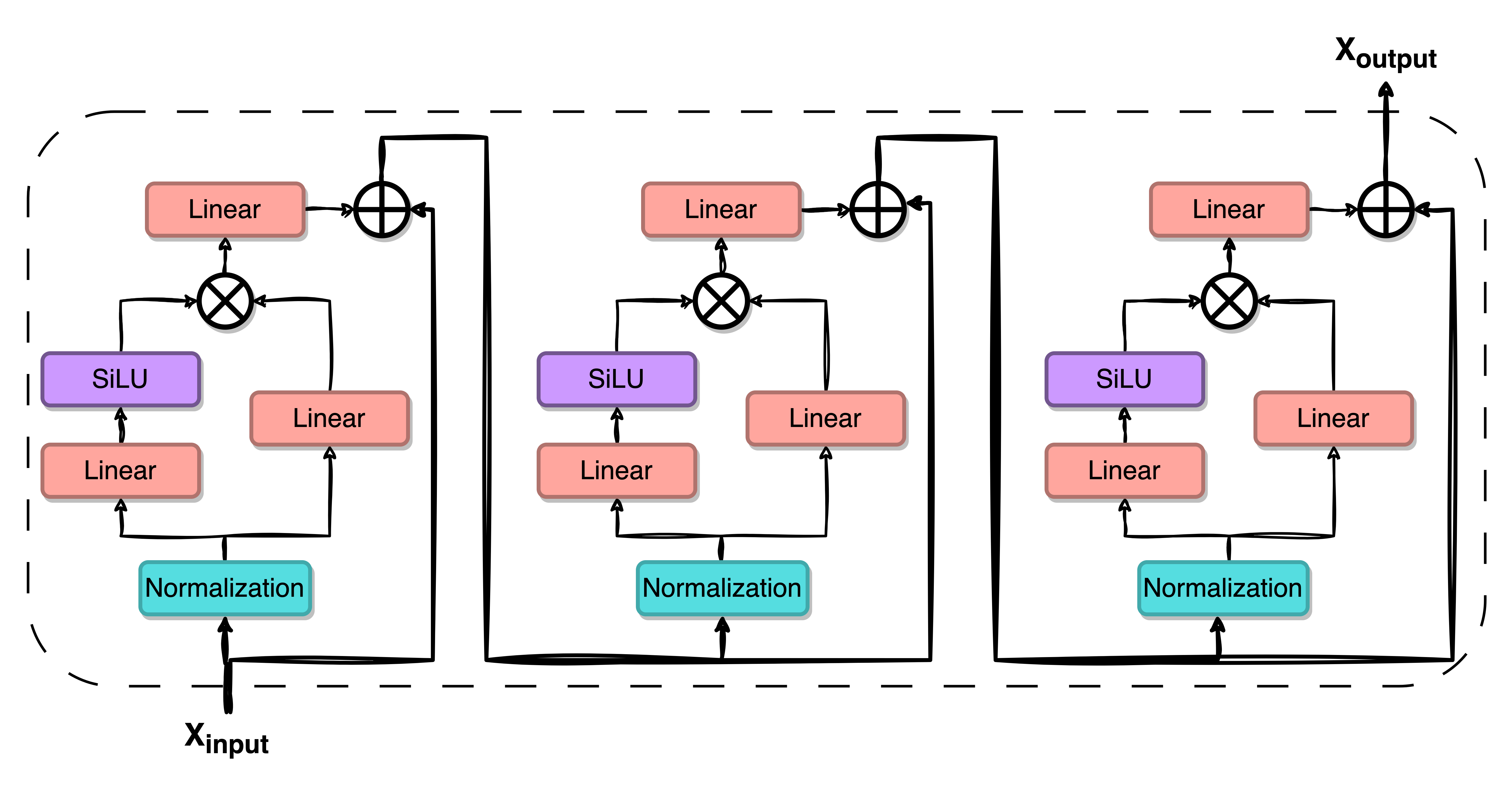}
    \caption{\textbf{A Detailed Structure of Multi-modal Understanding Adapter’s Dense Network.} Each sub-block consists of normalization, a linear layer, and SiLU activation function components. The input from the previous layer is also propagated to the next layer through a skip connection.}
    \label{fig:adapter-block}
\end{figure}
\vspace{-4mm}

\subsection{Multi-modal Understanding Adapters}

To integrate the feature embeddings into the LLaMA 2 model \cite{touvron2023llama}, we have developed multi-modal understanding adapters (see in Figure \ref{fig:model}) to align the output of multi-modal encoders with the input of the LLaMA 2 model. Different modal adapters share a unified architecture which includes a 1D convolutional layer, a linear projection layer, and a dense network composed of three sub-blocks, as depicted in Figure \ref{fig:adapter-block}. The process can be formulated as follows:


\vspace{-0.4cm}
\begin{gather*}
    \begin{aligned}
        X_{i} = X_{i-1} + L_{2,i}(SiLU(L_{1,i}(N_{i}(X_{i-1}))) \\ \times L_{3,i}(N_{i}(X_{i-1})))
    \end{aligned}
\end{gather*}

\noindent
where $X_i$ denotes the output embedding after the $i$-th sub-block, $L_{j,i}$ is the $j$-th linear layer in the $i$-th sub-block, and $N_i$ represents the normalization layer in the $i$-th sub-block. SiLU \cite{elfwing2018sigmoid} is the activation function. The architectural design of the adapter is inspired by the work of Liu et al. (2023) \cite{liu2023music} which demonstrated promising results in the field of music understanding. Subsequently, the output embeddings from the adapters, with a dimension of $4096$, serve as inputs for the LLaMA 2 model, providing multi-modal context information to enhance LLM's capabilities in music understanding, question answering, and guiding downstream music generation.



\subsection{LLM as a Bridge}


To introduce multi-modal context information to the LLM, we merge outputs from previous multi-modal understanding adapters into some specific hidden layers of the LLaMA 2 model. As shown in the light blue box of Figure \ref{fig:model}, the total number of hidden layers is $N=32$, and we introduce one modality-specific information every $L$-th layer ($L=6$) starting from the top (last) layer. For the lower $(N-3L-1)$ hidden layers, vanilla attention is employed, while the remaining layers above utilize zero-initialized attention. The input instruction prompt is fed into the first layer which is at the bottom of the Transformer, while the embedding vectors from music, image, and video are respectively imported into the $L$-th, $2L$-th, and $3L$-th layers starting from the top (last) layer. Through this approach, the LLaMA 2 model can effectively reason and comprehends multi-modal inputs.

\subsection{Music Understanding and Generation}



Inspired by models such as NExT-GPT \cite{wu2023next}, the M$^2$UGen model incorporates specialized audio tokens of the form $[AUD_i]$, where $i \in \{0,1,\cdots,7\}$, to distinguish between music question answering and generation tasks. The number of audio tokens is a hyper-parameter that determines the dimension of the input to the music output Transformer, also known as the output projector, during music generation. In the training phase, instruction sets containing music as the output are adjusted by appending such audio tokens to the end of the output. In the inference phase, the M$^2$UGen model will generate audio tokens only if the instruction prompt requires a music output. Through this method, the M$^2$UGen model exhibits the capability to produce text-only outputs for music question answering and music outputs for music generation within the same framework.

When performing the music generation task, we leverage the output from the output projector to modulate the music generation process. As each output token is mapped to a hidden embedding in the final layer of the LLaMA 2 model, we combine these hidden embeddings corresponding to the audio tokens with the audio token embeddings themselves as the input to the output projector. The subsequent output generated by the output projector serves as a crucial conditioning signal for the AudioLDM 2 \cite{liu2023audioldm2} / MusicGen \cite{copet2023simple} models, guiding the generation of the final output music. 

\subsection{Training Method}

Training a MLLM model from scratch is computationally expensive, which has led several models \cite{wu2023next,su2023pandagpt,li2023blip} to adopt the LoRA fine-tuning approach. In our training method, we alleviate the computational burden by freezing the encoders and generation models, focusing the training efforts on the multi-modal understanding adapters and output projector. This strategy not only reduces computational costs but also enhances training efficiency. To further simplify the training process and minimize the number of trainable parameters, we apply the LoRA method \cite{hu2022lora} to train the LLaMA 2 model. In the training pipeline, we utilize the following loss function to guide the optimization process:

\vspace{-4mm}
\[Loss = 
\begin{cases}
    L_{CE}(y_{tokens}, f(y)_{logits}) & \\
    + \Vert y_{embeddings} - g(f(x)_{hidden}) \Vert, & \text{if music} \\ \\
    L_{CE}(y_{tokens}, f(y)_{logits}), & \text{else}
\end{cases}\]

\noindent
where $y_{tokens}$ denotes target output tokens, $y_{embeddings}$ denotes the target embeddings for AudioLDM 2/MusicGen, $f(\cdot)$ represents the output from M$^2$UGen's LLaMA 2 model, $g(\cdot)$ represents the output from M$^2$UGen's output projection layer, and $L_{CE}$ is the cross entropy (CE) loss. The loss function used by the M$^{2}$UGen model is either CE alone or a combination of CE and mean squared error (MSE). During training, if the task involves only text token generation, the model is guided by the CE loss alone. If the task involves music generation, both CE and MSE are used together, with MSE calculated between the conditioning embedding generated by the output projector and the target music caption's text encoding from the music generation model. This dual-loss strategy ensures that the M$^{2}$UGen model is adept at both text token generation and generating embeddings for conditioning the downstream music generation models (AudioLDM 2 or MusicGen).

\section{Music Oriented Instruction Dataset}
\label{sec:mudataset}



Training MLLMs requires a significant amount of data. However, there is a noticeable deficiency of multi-modal datasets that specifically focus on music-related tasks. Currently, MusicCaps \cite{agostinelli2023musiclm} and MusicQA \cite{liu2023music} stand as the two largest publicly available dedicated datasets for music-related tasks, specifically in the areas of music captioning and music question answering, which are directly relevant to music understanding. Nevertheless, for the task of multi-modal music understanding and generation addressed in this paper, they are still far from sufficient. 

To train our M$^{2}$UGen model, the availability of multi-modal instruction datasets that support any-to-music generation is imperative. Moreover, extensive datasets that include pairs of different modalities, such as text-image pairs, are essential for modality alignment training. We make use of established datasets like Alpaca \cite{alpaca} with general knowledge for instruction following and COCO \cite{lin2014microsoft} for aligning the image encoder. In addition to utilizing existing resources, we also collect our own dataset. We adopt an automated approach to overcome the labor-intensive and time-consuming nature of manual data collection. Specifically, inspired by previous works \cite{liu2023music, gong2023listen}, the MU-LLaMA \cite{liu2023music} and MPT-7B \cite{MosaicML2023Introducing} models are utilized to generate required datasets. In the following subsections, we provide a comprehensive overview of the methodologies employed in crafting the datasets used for training the M$^{2}$UGen model.
 

\subsection{MUCaps Dataset}


We develop the MUCaps dataset which is composed of text-music pairs, encompassing approximately 1,200 hours of music sourced from AudioSet \cite{jort2017audioset} and some publicly accessible music websites. We leverage the MU-LLaMA model to caption the music files. For the captioning process, the MU-LLaMA model is presented with each music file along with the following question: \textit{``Describe the music in detail, including aspects such as instruments used, tempo, and the mood of the song''}. The generated MUCaps dataset is employed for the purpose of encoder and decoder alignment training.

\subsection{MUEdit Dataset}

To empower our model with the capability of performing music editing in response to prompts, we have curated the MUEdit dataset, which includes 55.69 hours of 10-second music pairs. The generation of this dataset is carried out through the following steps:

\begin{enumerate}[leftmargin=.2in]
    \setlength{\itemsep}{1.5mm}
    \item Generate captions for all the music files acquired using the MU-LLaMA model.
    
    \item Select pairs from a music pool, employing metrics such as tempo, beats, pitch, and magnitude to ensure that the chosen pairs exhibit similar rhythmic characteristics.

    \item For each selected pair, the MPT-7B model is employed to generate instructions. To create the human side of the conversation, the model is provided with the captions of the music files as input, accompanied by the following instructions:
    \begin{itemize}[leftmargin=.2in]
        \renewcommand\labelitemi{--}
        \setlength{\itemsep}{1mm}
        \item \textit{You are given description of input and target music}
        \item \textit{You will give a single line instruction of the form to change input music to target music}
        \item \textit{Give the instruction based on the music descriptions}
    \end{itemize}
    For generating the model side of the conversation, the model is supplied with the caption of the output music file, along with the following instructions:
    \begin{itemize}[leftmargin=.2in]
        \renewcommand\labelitemi{--}
        \setlength{\itemsep}{1mm}
        \item \textit{You are given description of a music}
        \item \textit{You will give a single line answer of the form `Here is a music that is ....'}
        \item \textit{Complete the answer based on the music description}
    \end{itemize}
\end{enumerate}

\subsection{MUImage Dataset}

We present the MUImage dataset for generating appropriate music for input images. The MUImage dataset is assembled by obtaining music samples from the AudioSet with paired videos. A random frame is selected from each video to serve as the input image. The process of creating this dataset involves the following steps:

\begin{enumerate}[leftmargin=.2in]
    \item Generate captions for all the music files acquired using the MU-LLaMA model.
    \item Generate captions for the corresponding images using the BLIP image captioning model \cite{li2022blip}.
    \item For each pair of music and image, the MPT-7B model is employed to generate instructions. The music and image captions are used as inputs. To create the human side of the conversation, the model is provided with the following instructions:
    \begin{itemize}[leftmargin=.2in]
        \renewcommand\labelitemi{--}
        \setlength{\itemsep}{1mm}
        \item \textit{You are given description of a music and an image}
        \item \textit{You will give a single line instruction of the form `Generate a music for the image that is ....' based on the image}
        \item \textit{Complete the instruction based on the music and image description}
    \end{itemize}
    For generating the model side of the conversation, the model is presented with the following instructions:
    \begin{itemize}[leftmargin=.2in]
        \renewcommand\labelitemi{--}
        \setlength{\itemsep}{1mm}
        \item \textit{You are given description of a music and an image}
        \item \textit{You will give a single line answer of the form `Here is a music that is ....' based on the image}
        \item \textit{Complete the answer based on the music and image description  }
    \end{itemize}
\end{enumerate}

\subsection{MUVideo Dataset}


Similar to MUImage, we also create the MUVideo dataset to enable our proposed M$^{2}$UGen framework to accomplish the task of video-to-music generation. The MUVideo dataset is curated by gathering music samples from the AudioSet as well with their corresponding videos. To construct this dataset, the following steps are designed:

\begin{enumerate}[leftmargin=.2in]
    \item Generate captions for all the music files acquired using the MU-LLaMA model.
    \item Generate captions for the corresponding videos using the VideoMAE captioning model \cite{tong2022videomae}.
    \item For each pair of music and video, the MPT-7B model is employed to generate instructions. The music and video captions are used as inputs. To create the human side of the conversation, the model is presented with the following instructions:
    \begin{itemize}[leftmargin=.2in]
        \renewcommand\labelitemi{--}
        \setlength{\itemsep}{1mm}
        \item \textit{You are given description of a music and a video}
        \item \textit{You will give a single line instruction of the form `Generate a music for the video that is ....'}
        \item \textit{Complete the instruction based on the music and video descriptions}
    \end{itemize}
    For generating the model side of the conversation, the model is given the following instructions:
    \begin{itemize}[leftmargin=.2in]
        \renewcommand\labelitemi{--}
        \setlength{\itemsep}{1mm}
        \item \textit{You are given description of a music and a video}
        \item \textit{You will give a single line answer of the form `Here is a music that is ....'}
        \item \textit{Complete the answer based on the music and video descriptions}
    \end{itemize}
\end{enumerate}

In these four datasets, we make efforts to minimize overlaps among the music files. Additionally, we establish evaluation splits for each dataset to facilitate the comparison of our model's performance with that of the current state-of-the-art (SOTA) models in their respective domains.
\section{Model Evaluation}


Given various capabilities demonstrated by our M$^{2}$UGen model, such as music understanding and music generation from multi-modal inputs, we conduct a comprehensive evaluation of the model in this section, assessing its performance across different subtasks. We also present a comparative analysis with other pertinent models. One such model demonstrating the capability of any-to-any generation is NExT-GPT\cite{wu2023next}; however, since the checkpoint released by the authors can not function as expected and has issues generating desired outputs, a direct comparison with NExT-GPT for large-scale evaluation is currently unfeasible. During the evaluation, we set the hyper-parameters of the M$^{2}$UGen model as follows: temperature$\ = 0.6$, top\_p$\ = 0.8$ and max target length$\ = 512$. We have also made sure that all models leveraging LLMs, such as LLaMA-Adapter \cite{gao2023llamaadapterv2} and SALMONN \cite{tang2023salmonn}, use the same hyper-parameters for evaluation to ensure a fair comparison. 

\subsection{Music Understanding}



To evaluate the music understanding capabilities of the M$^{2}$UGen model, we employ the MTG-eval-QA subset of the MusicQA dataset proposed by Liu et al. (2023) \cite{liu2023music} as our evaluation set, consisting of 4,500 music question-answer pairs. The SOTA models for comparison include LTU \cite{gong2023listen}, LLaMA-Adapter \cite{gao2023llamaadapterv2}, SALMONN \cite{tang2023salmonn} and MU-LLaMA \cite{liu2023music}. Among these models, MU-LLaMA stands out as the only one that was specifically trained on music-related datasets. The evaluation metrics used for music understanding follow the work of \cite{liu2023music}, containing BLEU (B-U) \cite{papineni2002bleu}, METEOR (M-R) \cite{banerjee2005meteor}, ROUGE$_L$ (R-L) \cite{lin2004rouge}, and BERT-Score (BERT-S) \cite{tianyi2020bertscore}.

\begin{table}[htbp]
\centering
\def\arraystretch{1.1}%
\caption{\textbf{Comparison of models for music understanding}. The best values of different metrics are made \textbf{bold}.}
\begin{tabular}{c|c|c|c|c}
\hline\hline
Model & \textbf{B-U$\uparrow$} & \textbf{M-R$\uparrow$} & \textbf{R-L$\uparrow$} & \textbf{BERT-S$\uparrow$} \\ \hline\hline
LTU & 0.242 & 0.274 & 0.326 & 0.887 \\ 
LLaMA Adapter & 0.273 & 0.334 & 0.413 & 0.895 \\ 
SALMONN & 0.286 & 0.332 & 0.371 & 0.898 \\
MU-LLaMA & 0.306 & 0.385 & 0.466 & 0.901 \\
\textbf{M$^2$UGen} & \textbf{0.308} & \textbf{0.393} & \textbf{0.476} & \textbf{0.902} \\ \hline\hline
\end{tabular}
\label{musicqa_eval}
\end{table}
\vspace{-2mm}

It is evident from the Table \ref{musicqa_eval} that the LTU, LLaMA Adapter, and SALMONN models are lagging behind MU-LLaMA and our M$^{2}$UGen model in the four metrics compared, since the former three models have not been specifically trained on music-related datasets. Both the MU-LLaMA and our M$^{2}$UGen models have been trained on the MusicQA dataset \cite{liu2023music}, demonstrating impressive music understanding capabilities. It is worth noting that our model still outperforms the MU-LLaMA model. One possible reason is that, in addition to the MusicQA dataset, we also have the MUCaps dataset for text-music alignment training, which contributes to enhancing our model's music understanding abilities. However, since the MUCaps dataset is generated by the MU-LLaMA model, there may be limitations on further improving our model's music understanding capabilities. This could be one of the factors preventing our model from significantly surpassing MU-LLaMA, which implies that in our future work, we may need to explore other music datasets to make further improvements.



\subsection{Text to Music Generation}

For text-to-music generation, we use the evaluation set from the MUCaps dataset. This set comprises 5,000 text-music pairs. SOTA models selected for comparison include CoDi \cite{tang2023any}, AudioLDM 2 \cite{liu2023audioldm2}, and MusicGen \cite{copet2023simple}. Among these models, MusicGen is the sole one explicitly trained for music generation, leading us to anticipate its superior performance compared to other models. In this and the following experiments, we evaluate the M$^2$UGen model with both AudioLDM 2 (M$^2$UGen v1) \cite{liu2023audioldm2} and MusicGen (M$^2$UGen v2) \cite{copet2023simple} decoders. Following MusicGen, we use the Fr{\'e}chet Audio Distance (FAD) \cite{kilgour2019FrchetAD}, Kullback-Leibler divergence (KL), and CLAP score \cite{laion2023clap} as the evaluation metrics.

\begin{table}[htbp]
\centering
\def\arraystretch{1.1}%
\caption{\textbf{Comparison of models for text-to-music generation}. The best values of different metrics are made \textbf{bold}.}
\begin{tabular}{c|c|c|c}
\hline\hline
Model & \textbf{FAD$_{vgg}$$\downarrow$} & \textbf{KL$\downarrow$} & \textbf{CLAP$_{score}$$\uparrow$} \\ \hline\hline
CoDi & 16.201 & 6.021 & 0.143 \\
AudioLDM 2 & 11.619 & 4.074 & 0.238 \\  
MusicGen & 10.697 & 3.909 & 0.289 \\
M$^{2}$UGen v1 & 11.143 & 3.982 & 0.282 \\
\textbf{M$^{2}$UGen v2} & \textbf{10.498} & \textbf{3.769} & \textbf{0.313} \\ \hline\hline
\end{tabular}
\label{text2musicgen_eval}
\end{table}
\vspace{-2mm}


From Table \ref{text2musicgen_eval}, we can see that CoDi performs the worst of all the models for the task of text-to-music generation. This is likely due to the fact that it has not been trained on music data. As mentioned above, our model incorporates AudioLDM 2 and MusicGen as music decoders, so in Table \ref{text2musicgen_eval}, M$^2$UGen v1 is compared to AudioLDM 2, and M$^2$UGen v2 is compared to MusicGen. It can be observed that our M$^2$UGen model performs better when given AudioLDM 2 or MusicGen as the music decoder compared to using them alone. Particularly, the generated music is more relevant to the input instructions, as indicated by the improvement in CLAP score. This can be attributed to the use of LLMs, which makes it easier for the model to understand the input instructions and use them to guide music generation.

\subsection{Prompt Based Music Editing}



M$^{2}$UGen is one of the few existing models that support music editing using natural language. In contrast, models like AUDIT \cite{wang2023audit} and InstructME \cite{han2023instructme} require specific prompt words such as ``Add'', ``Drop'', ``Remove'' and others to edit music. Loop Copilot \cite{zhang2023loop} is another model that supports music editing using natural language. However, since the authors have not made their model open-source, we are unable to compare it. Both AUDIT and InstructME also lack open-sourced models, but InstructME provides a few samples that we can utilize for comparison. Following AUDIT, in addition to using FAD and KL for evaluation, we introduce another metric called log spectral distance (LSD) to assess the music editing subtask.


\vspace{-0.1cm}
\begin{table}[htbp]
\centering
\def\arraystretch{1.1}%
\caption{\textbf{Comparison of models for prompt based music editing}. The best values of different metrics are made \textbf{bold}.}
\begin{tabular}{c|c|c|c}
\hline\hline
Model & \textbf{FAD$_{vgg}$$\downarrow$} & \textbf{KL$\downarrow$} & \textbf{LSD$\downarrow$} \\ \hline\hline
AUDIT & 2.855 & 6.267 & 0.987 \\ 
InstructME & 2.442 & 6.018 & 0.846 \\ 
M$^{2}$UGen v1 & 2.223 & 5.654 & 0.790 \\
\textbf{M$^{2}$UGen v2} & \textbf{2.191} & \textbf{5.118} & \textbf{0.735} \\ \hline\hline
\end{tabular}
\label{musicedit_eval}
\end{table}
\vspace{-2mm}

Table \ref{musicedit_eval} illustrates the superior performance of our M$^2$UGen model compared to the AUDIT and InstructME models. This achievement can be attributed to the utilization of the LLaMA 2 model for comprehending prompts in the editing task. This allows the model to effectively edit music based on natural language prompts. Furthermore, the use of the MERT Encoder enhances the model's capability to better understand the source music, consequently improving its effectiveness during the editing process.




\vspace{-0.1cm}
\begin{table}[htbp]
\centering
\def\arraystretch{1.1}%
\caption{\textbf{Comparison of models for image-to-music generation}. The best values of different metrics are made \textbf{bold}.}
\begin{tabular}{c|c|c|c}
\hline\hline
Model & \textbf{FAD$_{vgg}$$\downarrow$} & \textbf{KL$\downarrow$} & \textbf{IB Rank$\uparrow$} \\ \hline\hline
CoDi & 10.788 & 9.925 & 0.493 \\ 
M$^{2}$UGen v1 & 7.326 & 6.014 & 0.688 \\
\textbf{M$^{2}$UGen v2} & \textbf{6.968} & \textbf{5.878} & \textbf{0.819} \\ \hline\hline
\end{tabular}
\label{image2musicgen_eval}
\end{table}
\vspace{-0.4cm}

\begin{table}[htbp]
\centering
\def\arraystretch{1.1}%
\caption{\textbf{Comparison of models for video-to-music generation}. The best values of different metrics are made \textbf{bold}.}
\begin{tabular}{c|c|c|c}
\hline\hline
Model & \textbf{FAD$_{vgg}$$\downarrow$} & \textbf{KL$\downarrow$} & \textbf{IB Rank$\uparrow$} \\ \hline\hline
CoDi & 11.273 & 6.267 & 0.212 \\ 
CMT & 9.021 & 5.991 &  0.629 \\
M$^{2}$UGen v1 & 8.171 & 5.284 & 0.721 \\
\textbf{M$^{2}$UGen v2} & \textbf{8.002} & \textbf{4.939} & \textbf{0.850} \\\hline\hline
\end{tabular}
\label{video2musicgen_eval}
\end{table}

\subsection{Multi-modal Music Generation}


Multi-modal music generation from images/videos is a crucial ability of our M$^{2}$UGen model. CoDi \cite{tang2023any} is an any-to-any generation model, thus it is involved for both image-to-music (I2M) and video-to-music (V2M) generation experiments (Tables \ref{image2musicgen_eval} and \ref{video2musicgen_eval}), while CMT \cite{di2021video} is for video-to-music generation. The evaluation sets for I2M and V2M consist of 2,500 pairs of image-music and video-music, respectively. In these two sets of experiments, apart from FAD and KL, we introduce a new evaluation metric called ImageBind Ranking (IB Rank) \cite{girdhar2023imagebind} to assess the alignment between the image/video modality and the generated music. Specifically, we use the ImageBind model to obtain embeddings for the images/videos and the generated music, and then calculate their similarity scores for ranking purposes. 



From Tables \ref{image2musicgen_eval} and \ref{video2musicgen_eval}, it can be seen that our M$^{2}$UGen model demonstrates exceptional capabilities in multi-modal music generation, both in terms of the quality of generated music and the relevance to the input modality. Furthermore, it consistently outperforms other SOTA models. 

\subsection{Subjective Evaluation for Music Generation}

In order to provide a subjective assessment of our model's music generation capabilities, we conduct a subjective evaluation involving 40 participants. A total of 20 questions are created for three subtasks: text-to-music (T2M), image-to-music (I2M), and video-to-music (V2M) generation. Each question has options generated by the models to be compared, which are randomly shuffled to avoid any preference bias from the testers. Since the sample size for subjective evaluation is relatively small, we use the interactive demo released by the authors of NExT-GPT to generate evaluation results for the T2M and I2M subtasks. However, for the V2M experiment, the NExT-GPT demo occasionally fail to function properly, so we decide to omit its comparison for V2M. The results are presented in Table \ref{subjective_eval}. It shows that our proposed M$^{2}$UGen model consistently receive the highest preference among the testers for all three subtasks.

\vspace{-0.1cm}
\begin{table}[htbp]
\centering
\def\arraystretch{1.1}%
\caption{\textbf{Subjective comparison of models for music generation tasks}. The best values of different metrics are made \textbf{bold}.}
\begin{tabular}{c|c|c|c}
\hline\hline
Model & \textbf{T2M} & \textbf{I2M} & \textbf{V2M} \\ \hline\hline
CoDi & 14.75\% & 18.5\% & 17.5\% \\
AudioLDM 2 & 19.25\% & N/A & N/A \\  
MusicGen & 21.5\% & N/A & N/A \\
NExT-GPT & 15\% & 23.5\% & N/A \\
CMT & N/A & N/A & 37.5\% \\
\textbf{M$^{2}$UGen v2} & \textbf{29.5\%} & \textbf{58\%} & \textbf{45\%} \\ \hline\hline
\end{tabular}
\label{subjective_eval}
\end{table}
\vspace{-0.4cm}

\section{Conclusion and Future Work}

This paper introduces the M$^{2}$UGen model, which utilizes a large language model (LLM) to achieve music understanding and multi-modal music generation within a unified framework. Furthermore, we present a comprehensive methodology for generating the datasets used to train our model. The experiments show that our proposed M$^{2}$UGen model outperforms or achieves SOTA performance in various tasks, including music understanding, music editing, and text/image/video-to-music generation. Our future work will focus on further enhancing the model's fine-grained music understanding capabilities, as well as improving the correlation between generated music and input instructions. 


\typeout{}
{
    \small
    \bibliographystyle{ieeenat_fullname}
    \bibliography{main}
}

\begin{appendices}
The appendix presents supplementary details that extend beyond the content of the manuscript, aiming to enhance comprehension of the M$^2$UGen model. Comprehensive information is provided concerning the model's training dataset and training methodology, encompassing explicit insights into the utilized training approach and the corresponding model hyperparameters. Additionally, a thorough exposition is given regarding the composition of the evaluation sets employed in our study, accompanied by a delineation of the evaluation methodology and metrics applied to assess the performance of our model. To elucidate the diverse capabilities of our model, illustrative demo examples are also included.

\section{Music Oriented Dataset Information}

We generate 4 different datasets to train the M$^2$UGen model: MUCaps, MUImage, MUVideo and MUEdit datasets. The statistics of the datasets are given in Table \ref{data_stats}. An example of each from the 4 datasets are shown in Figure \ref{fig:dataset}.

\begin{table}[H]
\centering
\def\arraystretch{1.2}%
\caption{\textbf{Dataset Statistics}. The number of instructions in the dataset and total hours of music files in the dataset}
\begin{tabular}{c|c|c}
\hline \hline
\textbf{Dataset} & \textbf{Instruction Count} & \textbf{Hours of Music} \\ \hline \hline
\textbf{MUCaps}  & 21966                           & 1273.78                 \\ \hline
\textbf{MUImage} & 9966                            & 27.72                   \\ \hline
\textbf{MUVideo} & 13203                           & 36.72                   \\ \hline
\textbf{MUEdit}  & 10815                           & 60.22                   \\ \hline \hline
\end{tabular}
\label{data_stats}
\end{table}
\vspace{-4mm}

\begin{figure*}[t]
    \centering
    \includegraphics[width=\textwidth]{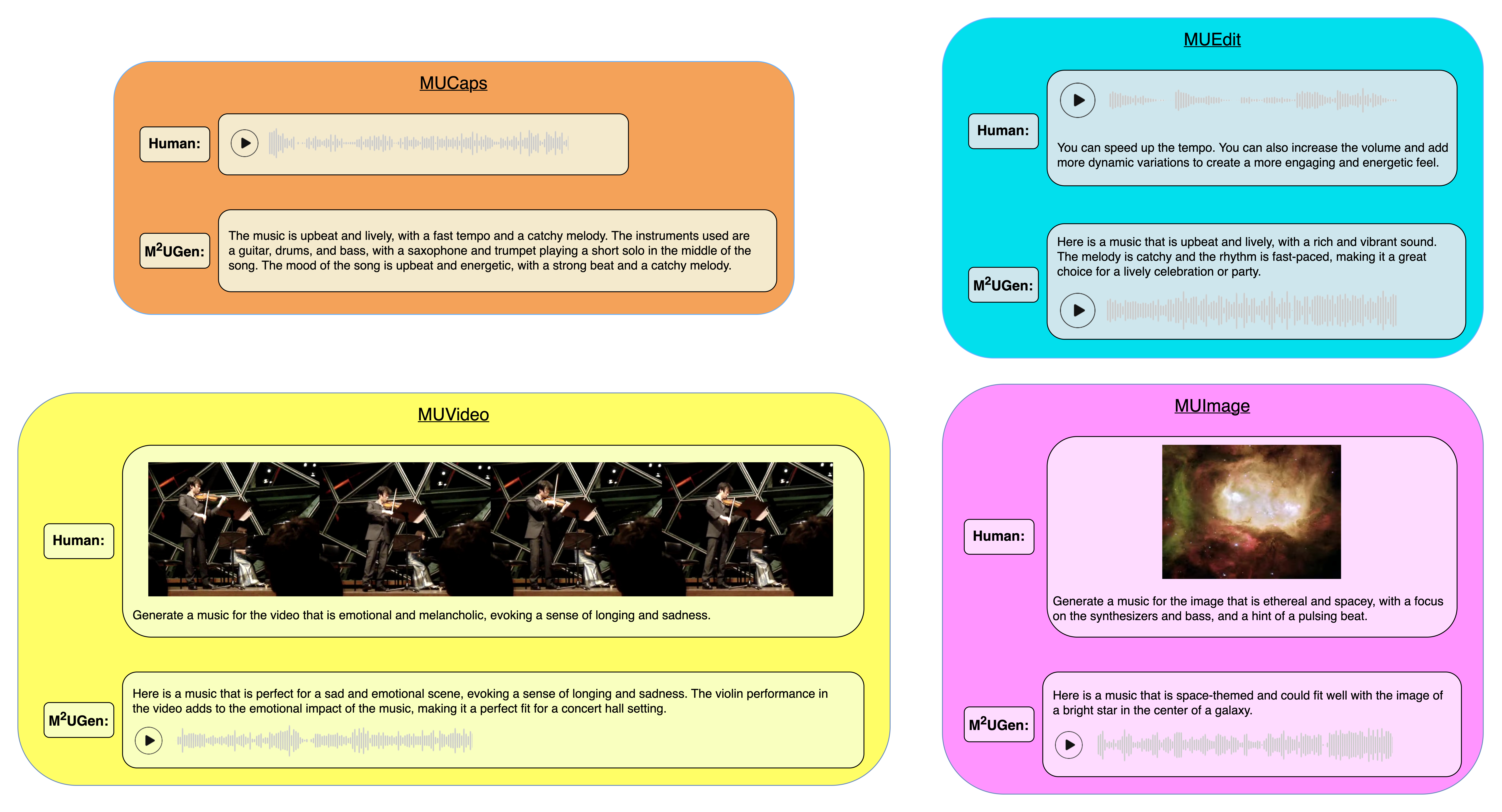}
    \caption{\textbf{Music Oriented Dataset.} Examples from the MUCaps, MUEdit, MUImage and MUVideo datasets used to train the M$^2$UGen model.}
    \label{fig:dataset}
\end{figure*}

\section{Model Training}

In this section, we detail the training strategy for the M$^2$UGen model along with parameters used for training.

\subsection{Model Training Strategy}

The M$^2$UGen model adopts the adapter training strategy, implementing a three-step training regimen. In the first phase, all parameters, with the exception of those associated with the Multi-modal Understanding Adapters, undergo freezing. The training dataset is configured to incorporate the MUCaps dataset for music understanding, the COCO dataset for image comprehension, and the captions sourced from the MUVideo dataset for video understanding. During this training stage, the Cross Entropy Loss function is applied to compute the disparity between the caption generated by the LLaMA 2 model and the target caption corresponding to the input modality. This process is illustrated in Figure \ref{fig:stage1}.

\begin{figure*}[t]
    \centering
    \includegraphics[width=\textwidth]{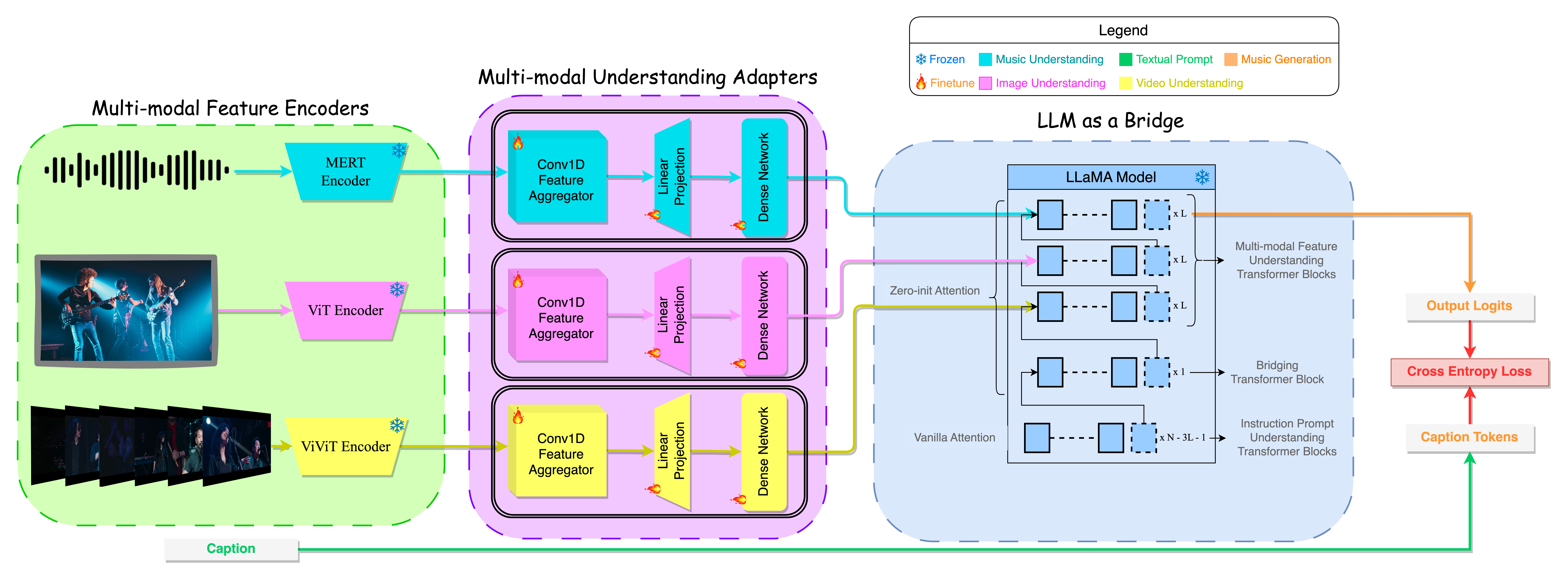}
    \caption{\textbf{Training Stage 1}: The Multi-modal Understanding Adapters are trained to integrate multi-modal features into the different layers of the LLaMA 2 model.}
    \label{fig:stage1}
\end{figure*}

In the second training stage, the output projector is trained to generate conditional embeddings using input captions processed by the LLaMA 2 model. The LLaMA 2 model produces specialized audio tokens, denoted as [AUD$i$] where $i \in \{1, 2, \ldots, K\}$ (with $K$ as a hyperparameter representing the number of special audio tokens added to the LLaMA 2 model's vocabulary) when processing input captions. The special audio tokens serve as signaling indicators, aiding the model in determining whether to generate text+music or solely text. In training, these audio tokens are added to the end of the text output in datasets requiring music output. During inference, if the M$^2$UGen model generates audio tokens, downstream music decoders (MusicGen/AudioLDM 2) will perform music generation, otherwise, solely text will be produced.

The hidden embeddings corresponding to these audio tokens from the last layer of the LLaMA 2 model is then input into the output projection layer, generating the conditional embedding for the Music Generation model. The MUCaps dataset is utilized to train this stage, with captions serving as inputs to the model and the target output tokens set as the special audio tokens.

\begin{figure*}[t]
    \centering
    \includegraphics[width=\textwidth]{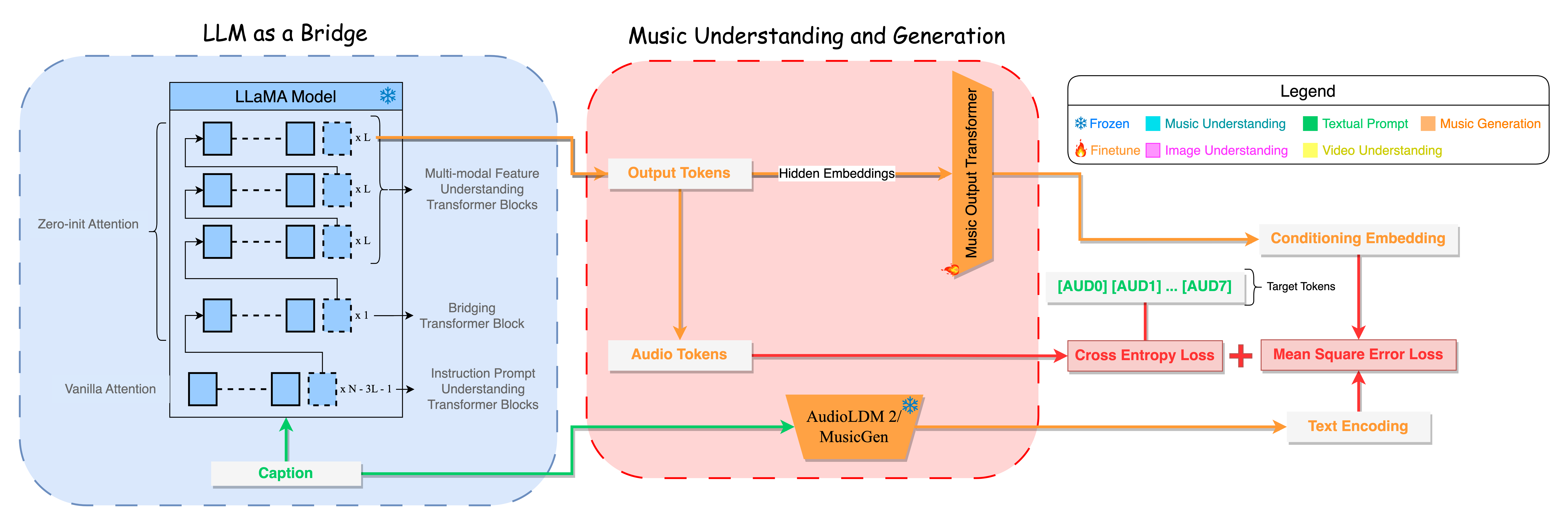}
    \caption{\textbf{Training Stage 2}: The Output Projection Layer is trained to generate the conditioning embedding for the MusicGen/AudioLDM 2 model.}
    \label{fig:stage2}
\end{figure*}

Assuming a total of $N$ tokens generated by the LLaMA 2 model, where [AUD$i$] with $i \in \{1, 2, \ldots, 8\}$ constitutes the last $8$ tokens. The hidden embeddings size is $(1, N, 4096)$, and the last 8 tokens are extracted along dimension $-1$, resulting in an input embedding size of the Output Projection layer as $(1, 8, 4096)$. The output size from the projection layer varies based on the Music Generation model: for AudioLDM2, it is $(1, 512)$, and for MusicGen, it is $(512, 768)$.

\vspace{2mm}

In the final training stage, the LoRA training strategy is employed to train the LLaMA 2 model, concurrently fine-tuning the Multi-modal Understanding Adapter and Output Projection layer. This stage utilizes datasets including Alpaca, MusicQA, MUImage, MUVideo, and MUEdit. To signal the M$^2$UGen model to generate both music and text, the output text in MUImage, MUVideo, and MUEdit datasets is extended with the special audio tokens.

\begin{figure*}[t]
    \centering
    \includegraphics[width=\textwidth]{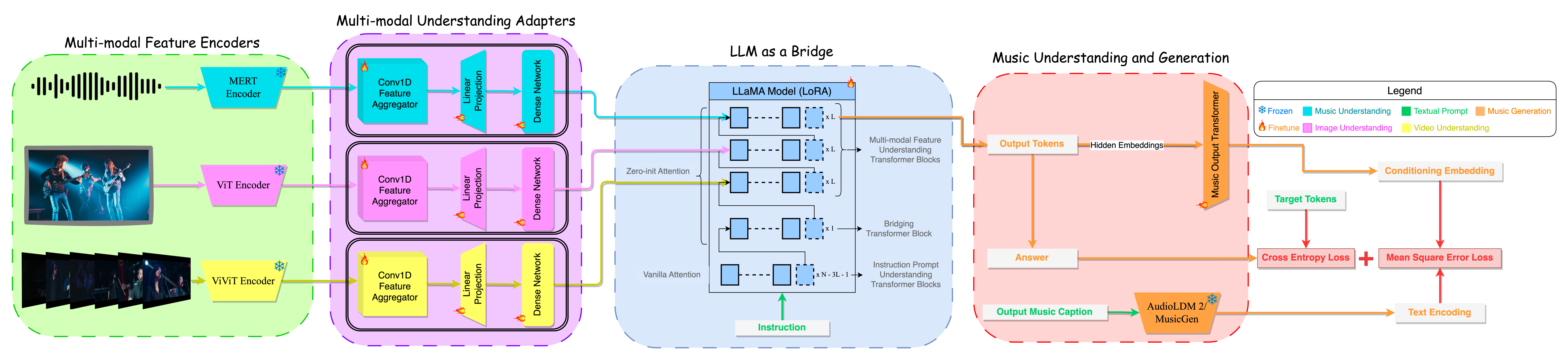}
    \caption{\textbf{Training Stage 3}: The Multi-modal Understanding Adapter and Output Projection Layer are fine-tuned while the LoRA-enabled LLaMA 2 model is trained in this stage.}
    \label{fig:stage3}
\end{figure*}

\subsection{Reasoning for Training Strategy}

Using the initial training stage, the model undergoes training with the objective of comprehending diverse modalities by leveraging extensive captioning datasets for music, image, and video. The subsequent training stage focuses on refining the LLaMA 2 model's capability to condition music generation based on input captions.

These dual training stages equip the model with the ability to grasp various modalities and generate distinct music conditions based on input captions. This proficiency significantly contributes to the final training stage. In this ultimate phase, the model leverages the trained Multi-modal Understanding Adapters and Output Projection layers to bridge the gap between them, honing the LLaMA 2 model's skills through the utilization of multi-modal music generation and music understanding datasets.

\subsection{Model Training Parameters}

We conduct training for the three stages of our model, employing 5, 5, and 2 epochs, respectively. The training process incorporates the following hyper-parameters: $N=32$, $L=6$, $\text{number of Audio Tokens}=8$, and $lr=10^{-4}$. This choice of hyper-parameters, coupled with our training strategy, allows for the effective use of a reduced number of epochs and a smaller dataset during the final stage of model training.

\section{Model Evaluation}

In this section, we elaborate on the datasets employed to assess the various capabilities of the M$^2$UGen model, followed by a discussion of the evaluation metrics utilized.

\subsection{Evaluation Datasets}

For each of the training datasets generated—MUCaps, MUImage, MUVideo, and MUEdit—we create a corresponding evaluation set. The methodology employed for generating the evaluation dataset mirrors that of the training dataset generation. Detailed statistics for the evaluation datasets are provided in Table \ref{eval_stats}. However, for the MUEdit dataset, the evaluation set could not be utilized for model evaluation due to the unavailability of code bases and trained checkpoints for InstructME\cite{han2023instructme} and AUDIT\cite{wang2023audit}. Consequently, we resort to utilizing samples from InstructME's demo website, which includes samples from both AUDIT and InstructME, to assess our model's performance. For evaluating the M$^2$UGen's music understanding capabilities we utilize the evaluation split of the MusicQA dataset.

\begin{table}[H]
\centering
\def\arraystretch{1.2}%
\caption{\textbf{Evaluation Dataset Statistics}. The number of instructions in the evaluation dataset and total hours of music files in the dataset}
\begin{tabular}{c|c|c}
\hline \hline
\textbf{Dataset} & \textbf{Instruction Count} & \textbf{Hours of Music} \\ \hline \hline
\textbf{MUCaps}  & 4000                           & 265.35                 \\ \hline
\textbf{MUImage} & 2500                            & 6.94                   \\ \hline
\textbf{MUVideo} & 2500                           & 6.94                   \\ \hline
\textbf{MUEdite}  & 2000                           & 5.55                   \\ \hline \hline
\end{tabular}
\label{eval_stats}
\end{table}

\subsection{Evaluation Metrics}

To assess music question answering, we adopt the metrics employed in \cite{liu2023music}, namely BLEU (B-U) \cite{papineni-etal-2002-bleu}, METEOR (M-R) \cite{banerjee-lavie-2005-meteor}, ROUGE$_L$ (R-L) \cite{lin-2004-rouge}, and BERT-Score (BERT-S) \cite{bert-score}. These metrics are widely used for evaluating text generation. For all music generation tasks, we employ Fr{\'e}chet Audio Distance (FAD) \cite{kilgour2019FrchetAD} and Kullback-Leibler divergence (KL), as these metrics are commonly utilized to assess the quality of generated audio. In addition to these general metrics, task-specific metrics are applied for each of the music generation tasks, namely Text-to-Music, Image-to-Music, Video-to-Music, and Music Editing.

In the context of Text-to-Music, we employ the CLAP\cite{laion2023clap} score, calculated by determining the cosine similarity between the CLAP embedding for the generated music and the text input:
\[
    CLAPScore(M, T) = \max(100 \times \cos(E_M, E_T), 0)
\]
Here, $M$ represents the generated music, $T$ denotes the text input, and $E_M$, $E_T$ represent the CLAP embeddings for the music and text, respectively.

For the Music Editing task, we leverage the Log Spectral Distance (LSD) to assess the disparity between the generated music and the target music. This metric facilitates the evaluation of whether the frequencies in the generated music, post-editing, align with those in the target music.

For Image-to-Music and Video-to-Music tasks, we introduce the ImageBind\cite{girdhar2023imagebind} Ranking (IB Rank), akin to the CLAP score, to quantify the alignment between the input modality and the generated music. Considering $N$ distinct models producing $N$ music files, we generate ImageBind embeddings for the music files, denoted as $E_{M1}, E_{M2}, \ldots, E_{MN}$, as well as the ImageBind embedding for the input modality, denoted as $E_{I/V}$. The embeddings for the music files are ranked based on their cosine similarity to $E_{I/V}$. Subsequently, after ranking all the samples in the evaluation set, the ImageBind Ranking is computed by calculating the ranking score using the individual rankings.

Using these evaluation metrics, we are able to evaluate the M$^2$UGen model against other state-of-the-art models for the different tasks.

\section{Model Demonstration}

In this section, we present screenshots of the M$^2$UGen model demo, illustrating various capabilities of the model.

Figures \ref{fig:T2M+MU}, \ref{fig:I2M+MU}, and \ref{fig:V2M+MU} showcase the M$^2$UGen model's ability to generate music directly from textual prompts and draw inspiration from images and videos, both with and without textual guidance. Figure \ref{fig:ME+MU} exemplifies M$^2$UGen's proficiency in music editing guided by natural language prompts. Additionally, Figures \ref{fig:T2M+MU+ME}, \ref{fig:I2M+MU+ME}, and \ref{fig:V2M+MU+ME} illustrate the utilization of M$^2$UGen's editing capabilities to further refine music generated from different modalities. Collectively, the M$^2$UGen model proves to be a robust framework for Music Understanding, Generation, and Editing.

\raggedbottom

\begin{figure}[H]
    \centering
    \includegraphics[width=0.9\columnwidth]{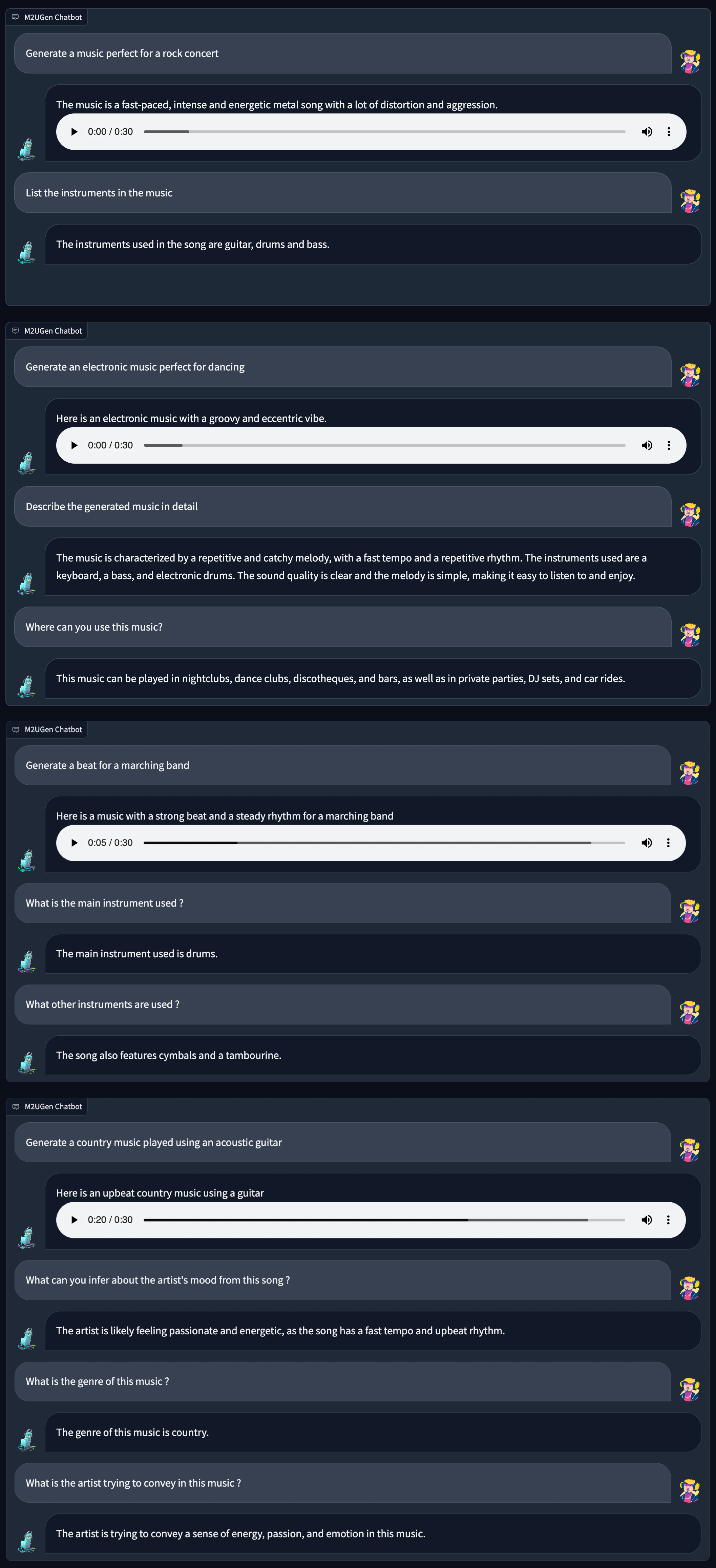}
    \caption{\textbf{Text-To-Music Generation and Understanding:} The M$^2$UGen model is capable of generating music from text prompts and also answering questions regarding the generated music.}
    \label{fig:T2M+MU}
\end{figure}

\begin{figure}[t]
    \centering
    \includegraphics[width=0.9\columnwidth]{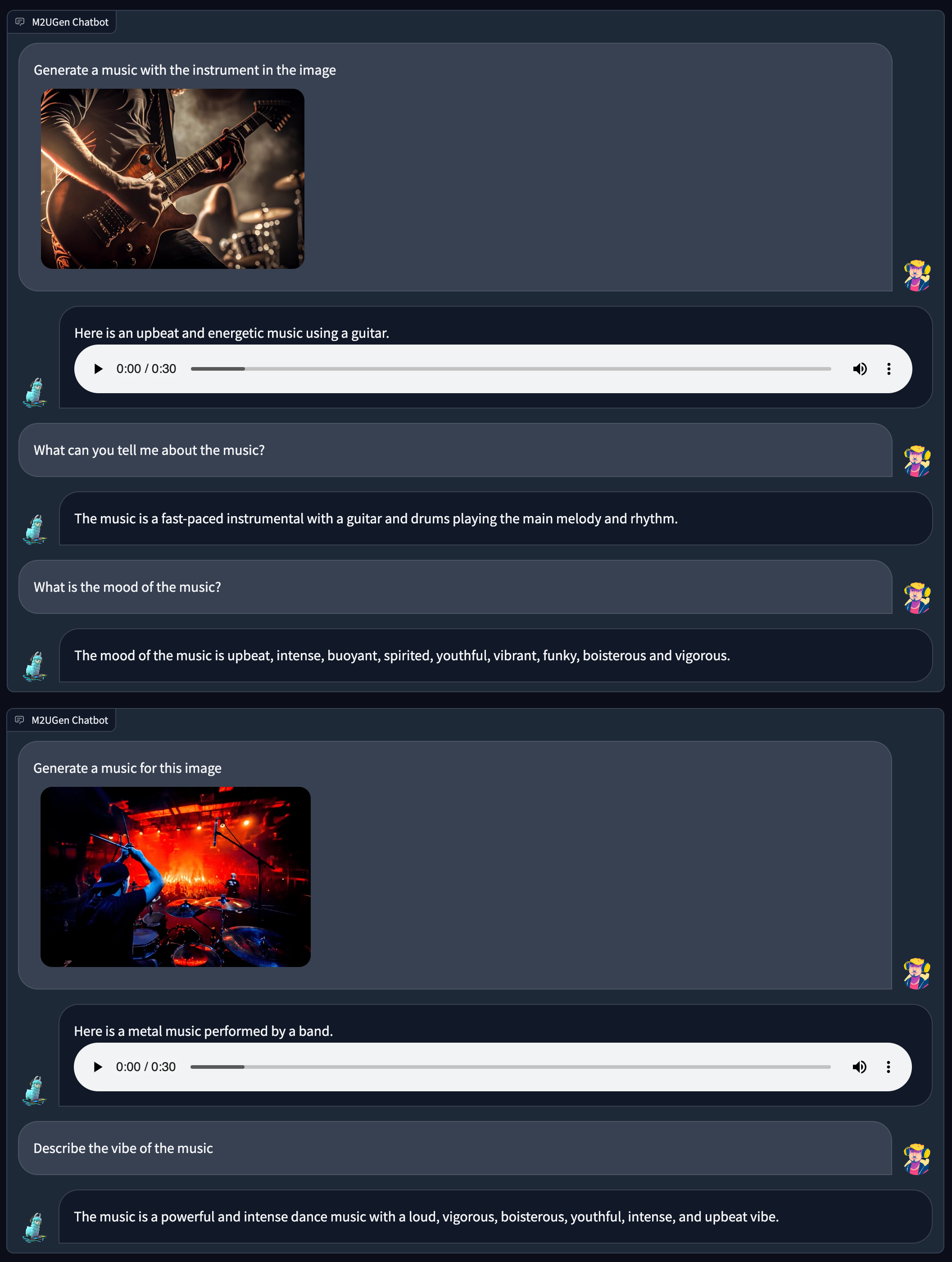}
    \caption{\textbf{Image-To-Music Generation and Understanding:} The M$^2$UGen model is capable of generating music for images and also answering questions regarding the generated music.}
    \label{fig:I2M+MU}
\end{figure}

\begin{figure}[t]
    \centering
    \includegraphics[width=0.9\columnwidth]{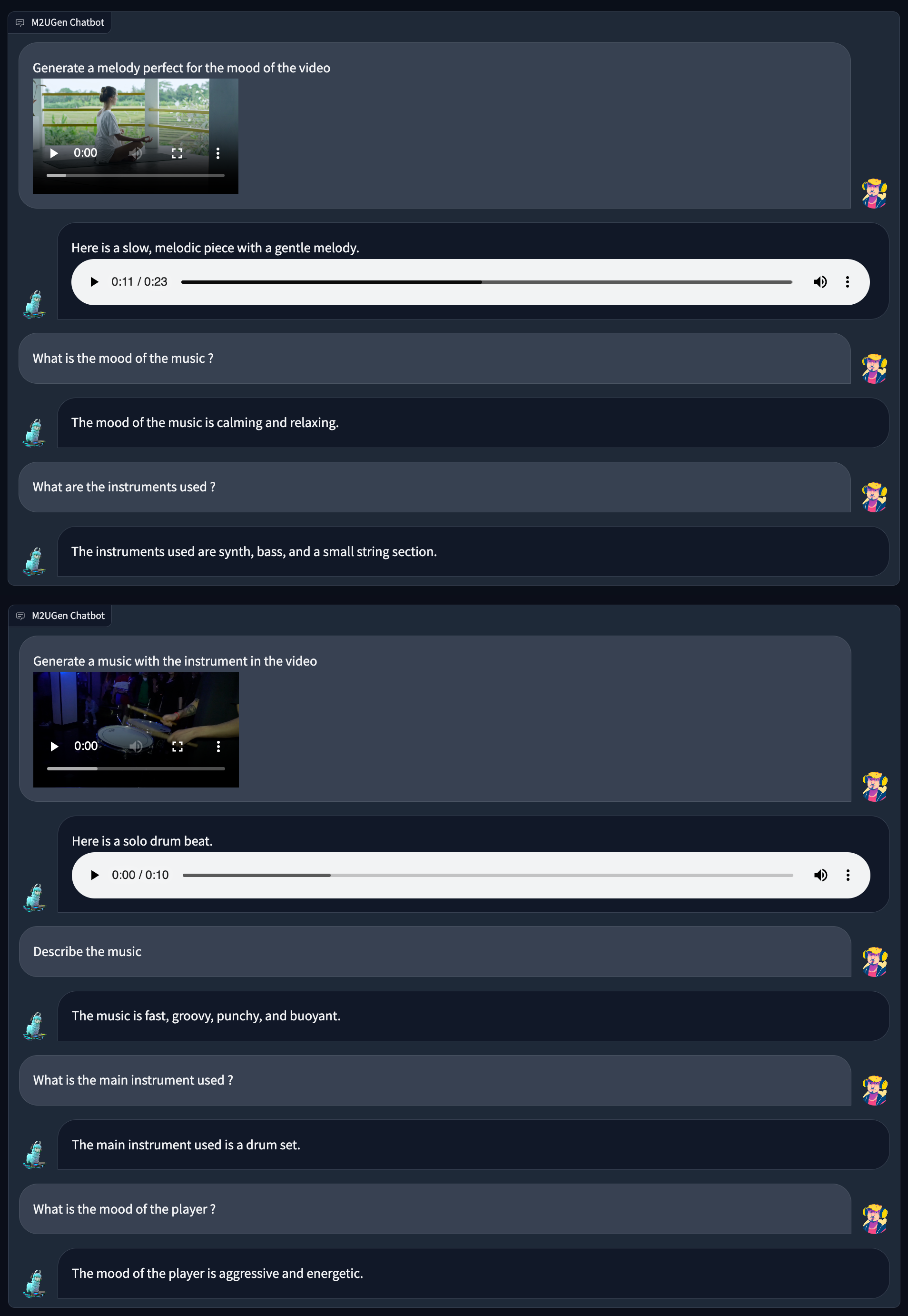}
    \caption{\textbf{Video-To-Music Generation and Understanding:} The M$^2$UGen model is capable of generating music for videos and also answering questions regarding the generated music.}
    \label{fig:V2M+MU}
\end{figure}

\begin{figure}[t]
    \centering
    \includegraphics[width=0.9\columnwidth]{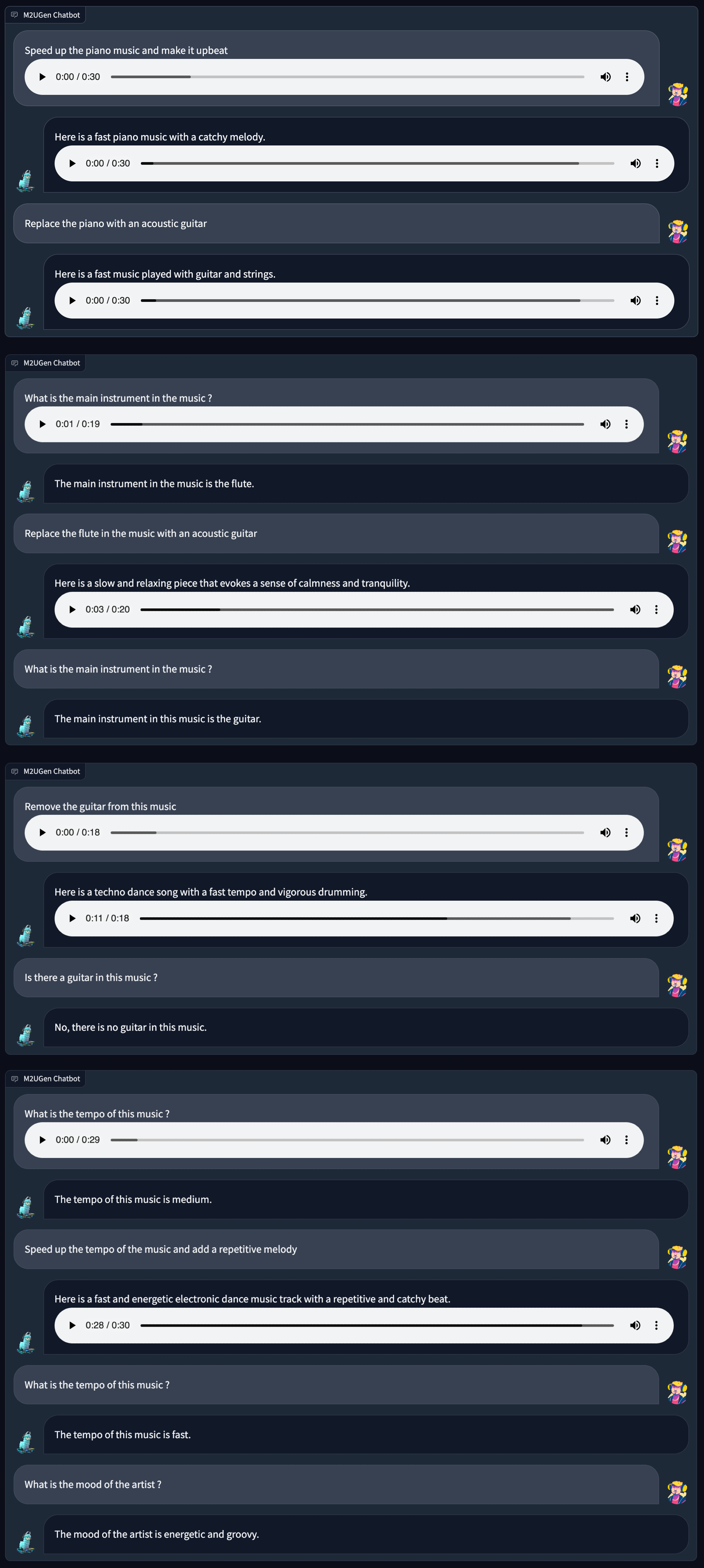}
    \caption{\textbf{Music Editing and Understanding:} The M$^2$UGen model is capable of editing input music based on natural language prompts and also answering questions regarding the generated music.}
    \label{fig:ME+MU}
\end{figure}

\begin{figure}[t]
    \centering
    \includegraphics[width=0.9\columnwidth]{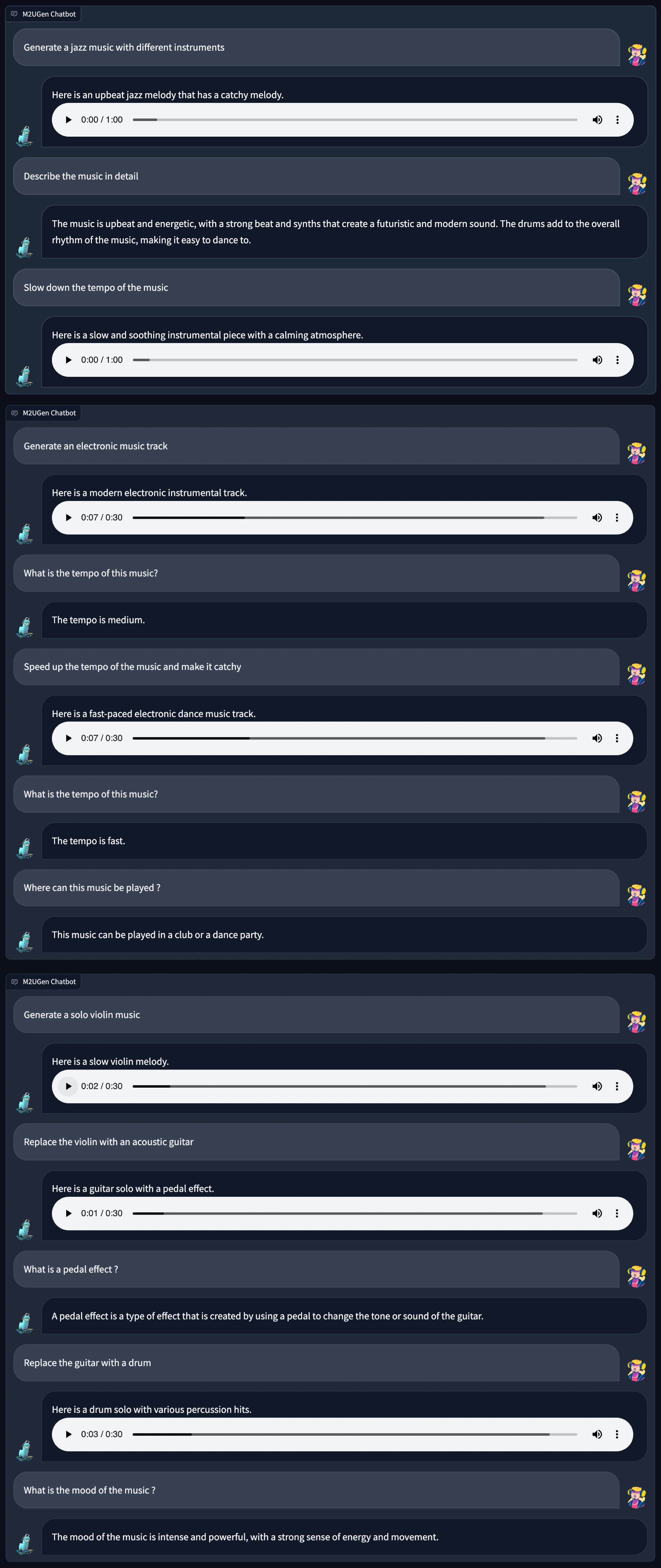}
    \caption{\textbf{Text-To-Music Generation and Understanding + Music Editing:} The M$^2$UGen model is capable of generating music from text prompts, answering questions regarding the generated music and also editing the generated music using Natural Language prompts.}
    \label{fig:T2M+MU+ME}
\end{figure}

\begin{figure}[t]
    \centering
    \includegraphics[width=0.9\columnwidth]{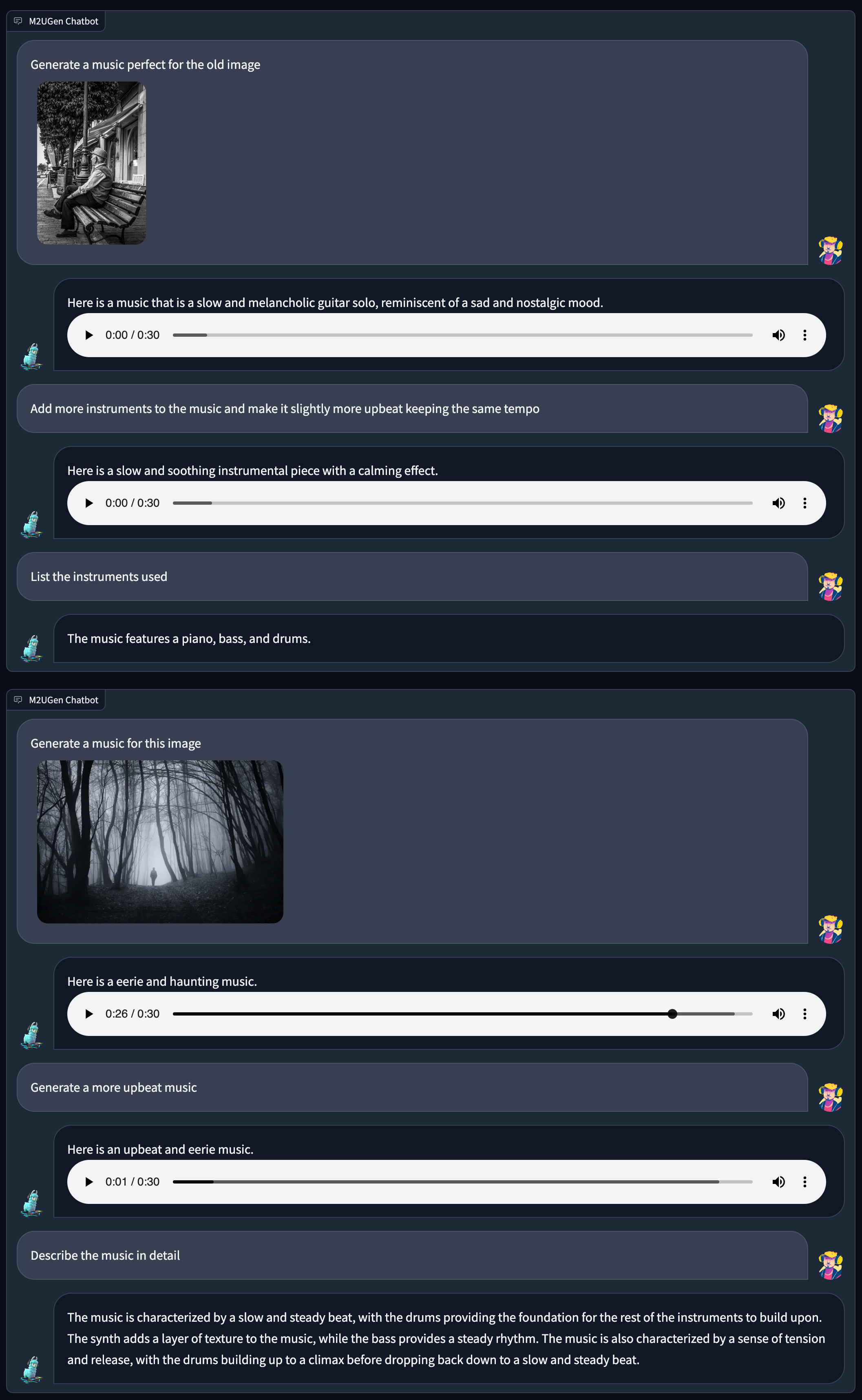}
    \caption{\textbf{Image-To-Music Generation and Understanding + Music Editing:} The M$^2$UGen model is capable of generating music for images, answering questions regarding the generated music and also editing the generated music using Natural Language prompts.}
    \label{fig:I2M+MU+ME}
\end{figure}

\begin{figure}[t]
    \centering
    \includegraphics[width=0.9\columnwidth]{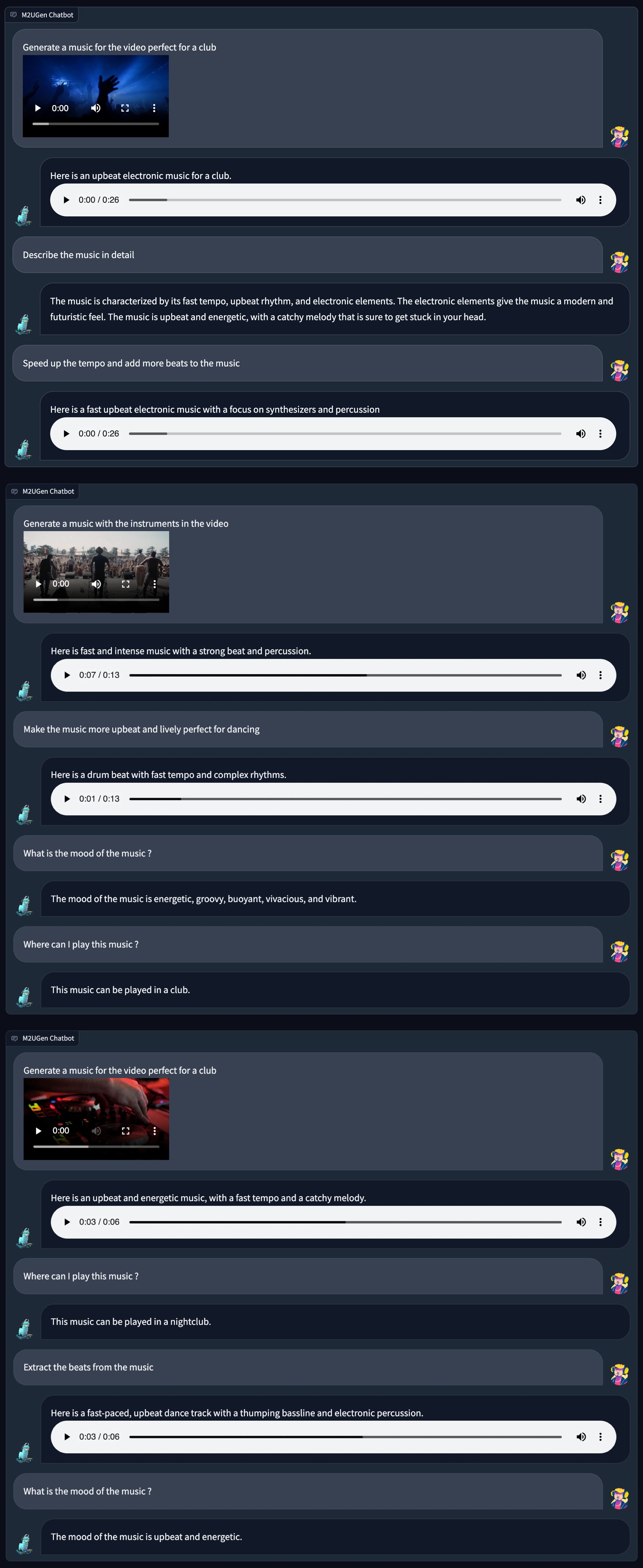}
    \caption{\textbf{Video-To-Music Generation and Understanding + Music Editing:} The M$^2$UGen model is capable of generating music for videos, answering questions regarding the generated music and also editing the generated music using Natural Language prompts.}
    \label{fig:V2M+MU+ME}
\end{figure}

\end{appendices}

\end{document}